\documentclass[Physsubmission, Phys]{SciPost}
\hypersetup{colorlinks,
linkcolor={red!50!black},citecolor={blue!50!black},urlcolor={blue!80!black}}
\usepackage[bitstream-charter]{mathdesign}
\DeclareSymbolFont{usualmathcal}{OMS}{cmsy}{m}{n}
\DeclareSymbolFontAlphabet{\mathcal}{usualmathcal}
\usepackage{fixmath}
\usepackage{graphicx,wrapfig}
\usepackage{color}

\begin{document}

\begin{center}{\Large \textbf{
Possible studies of gluon transversity in the spin-1 deuteron\\
at hadron-accelerator facilities\\
}}\end{center}

\begin{center}
S. Kumano\textsuperscript{1,2$\star$} and
Qin-Tao Song\textsuperscript{3} 
\end{center}

\begin{center}
{\bf 1} KEK Theory Center,
             Institute of Particle and Nuclear Studies, KEK,\\
             Oho 1-1, Tsukuba, Ibaraki, 305-0801, Japan
\\
{\bf 2} J-PARC Branch, KEK Theory Center,
             Institute of Particle and Nuclear Studies, KEK, \\
           and Theory Group, Particle and Nuclear Physics Division, 
           J-PARC Center, \\
           Shirakata 203-1, Tokai, Ibaraki, 319-1106, Japan
\\
{\bf 3} School of Physics and Microelectronics, Zhengzhou University, \\
             Zhengzhou, Henan 450001, China
\\
* shunzo.kumano@kek.jp, songqintao@zzu.edu.cn
\end{center}

\begin{center}
\today
\end{center}

%%%%%%%%%%%%%%%%%%%%%%%%%%%%%%%%%%%%%%%%
\definecolor{palegray}{gray}{0.95}
\begin{center}
\colorbox{palegray}{
  \begin{tabular}{rr}
  \begin{minipage}{0.1\textwidth}
    \includegraphics[width=22mm]{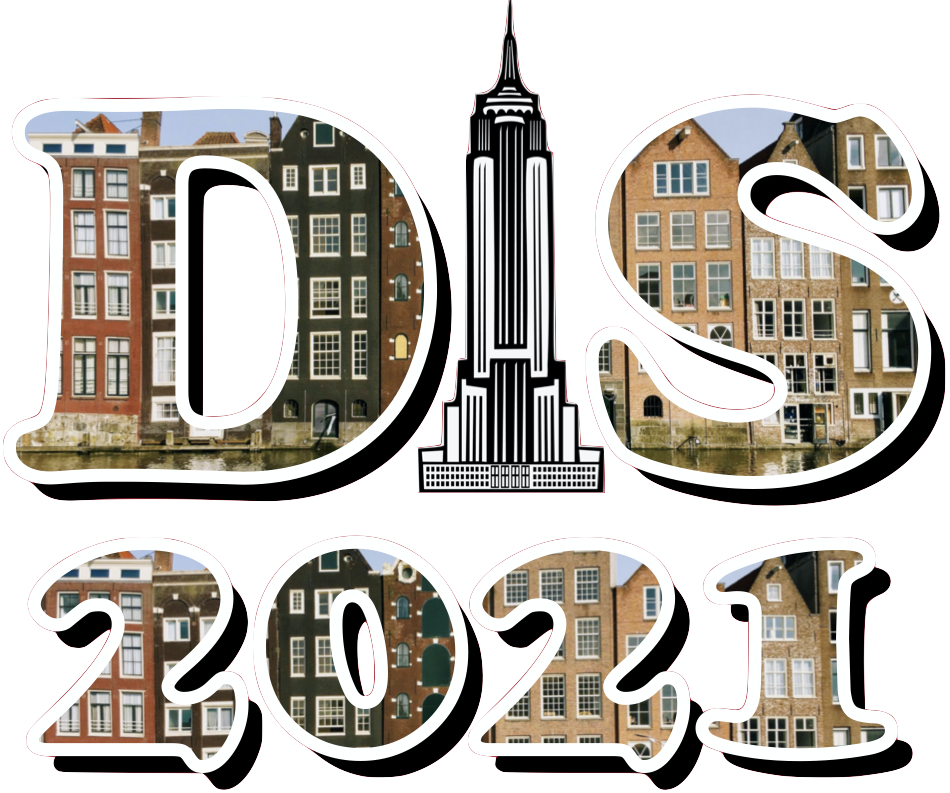}
  \end{minipage}
  &
  \begin{minipage}{0.75\textwidth}
    \begin{center}
    {\it Proceedings for the XXVIII International Workshop\\ on Deep-Inelastic Scattering and
Related Subjects,}\\
    {\it Stony Brook University, New York, USA, 12-16 April 2021} \\
    \doi{10.21468/SciPostPhysProc.?}\\
    \end{center}
  \end{minipage}
\end{tabular}
}
\end{center}
%%%%%%%%%%%%%%%%%%%%%%%%%%%%%%%%%%%%%%%%

\section*{Abstract}
{\bf
Chiral-odd gluon transversity distribution
could shed light on a new aspect of hadron physics. 
Although we had much progress recently on 
quark transversity distributions, there is no experimental measurement
on the gluon transversity. The gluon trasversity does not exist 
in the spin-1/2 nucleons and it exists in the spin-1 deuteron.
Therefore, it could probe new hadron physics in the deuteron 
beyond the basic bound system of a proton and a neutron
because the nucleons cannot contribute directly.
Here, we explain that the gluon transversity can be measured 
at hadron accelerator facilities, such as Fermilab and NICA, 
in addition to charged-lepton scattering measurements 
at lepton accelerator facilities by showing cross sections
of the proton-deuteron Drell-Yan process as an example.
}

\vspace{-0.40cm}
%%%%%%%%%%%%%%%%%%%%%%%%%%%%%%%%%%%%%%%%%%%%%%%%%%%%%%%%%%%%%%%%%%%%%%%%%%%%%%%%
\section{Introduction}
\label{intro}
\vspace{-0.20cm}

There are tensor-polarized structure functions for spin-1 hadrons,
such as the deuteron, in addition to the polarized ones 
for the spin-1/2 nucleons. 
The twist-2 tensor-polarized structure function $b_1$ was measured
by the HERMES collaboration \cite{Airapetian:2005cb}.
We have been investigating structure functions of spin-1 hadrons
theoretically \cite{our-studies,ks-trans-g-2020}.
On the other hand, there are future experimental projects 
to measure these new structure functions at various accelerator facilities. 
The structure function $b_1$ will be measured at the Jefferson 
laboratory (JLab) \cite{jlab-b1}, 
and there is a letter of intent to measure 
the gluon transversity also at JLab \cite{jlab-gluon-trans}. 
These polarized-deuteron structure functions will
be also investigated at Fermilab \cite{Fermilab-dy}
and Nuclotron-based Ion Collider fAcility (NICA) \cite{nica}
with the polarized-deuteron target and beam, respectively.
Furthermore, they should be also investigated
at future electron-ion colliders in US and China.

Recently, there was much progress in the quark transversity 
distribution on both theoretical and experimental sides.
Global analyses of experimental data were done and optimum
quark transversity distributions were proposed.
However, there is no experimental measurement 
on the gluon transversity.
The gluon transversity is a chiral-odd distribution, which needs
a helicity-flip for a gluon in the gluon-hadron helicity amplitude.
Namely, the two-unit of spin $\Delta s=2$ is necessary, so that
the hadron spin should be equal to one or larger,
and it does not exist in the spin-1/2 nucleons.
Because of this property, the gluon transversity is a unique distribution 
for probing the non-nucleonic component in the deuteron, it could
provide a new hadron-physics information beyond a simple bound system 
of nucleons.

There is an experimental plan to measure the gluon transversity
at JLab by measuring the azimuthal angle dependence of the deuteron
spin polarization. Since there are a number of hadron accelerators
in operation and will be built in the near future, it is nice 
that gluon transversity experiments will be done as independent
and complementary measurements to the JLab one.
These projects were not possible until recently because there was
no theoretical formalism with the gluon transversity
for hadron-facility experiments. 
%%%%%
Considering this situation, we provided such a formalism 
for the gluon transversity studies at the hadron accelerator facilities 
by taking the proton-deuteron Drell-Yan process as an example 
\cite{ks-trans-g-2020}.
In the near future, we expect that the gluon transversity experiments
will be realized at Fermilab \cite{Fermilab-dy}
and NICA \cite{nica} by using proton and deuteron beams.

\vspace{-0.40cm}
%%%%%%%%%%%%%%%%%%%%%%%%%%%%%%%%%%%%%%%%%%%%%%%%%%%%%%%%%%%%%%%%%%%%%%%%%%%%%%%%
\section{Linear polarization of spin-1 deuteron}
\label{linear-polarization}
\vspace{-0.20cm}

The gluon transversity is defined by linear polarizations
for the spin-1 deuteron and also the gluon. 
Although the linear polarization is often used for photon,
it is rarely used for hadron polarizations, so that 
it is briefly explained in this section.
%%%
The spin-1 deuteron polarizations are described by 
the polarization vectors 
defined by
$\vec E_\pm = \left ( \, \mp 1,\, -i,\, 0 \, \right ) /\sqrt{2}$, 
$\vec E_0  = \left ( \, 0,\, 0,\, 1 \, \right )$,
$\vec E_x  = \left ( \, 1,\, 0,\, 0 \, \right )$, and
$\vec E_y  = \left ( \, 0,\, 1,\, 0 \, \right )$.
The polarizations $\vec E_+$, $\vec E_0$, and $\vec E_-$
indicate the spin states with the $z$ component of spin
$s_z =+1$, $0$, and $-1$; 
$\vec E_x$ and $\vec E_y$ are linear polarizations.
The spin vector $\vec S$ and tensor $T_{ij}$ 
are written by these polarization vectors, and then 
they are parametrized in the deuteron rest frame
as
\vspace{-0.10cm}
\begin{align}
\vec S
& = \text{Im} \, (\, \vec E^{\, *} \times \vec E \,)
= (S_{T}^x,\, S_{T}^y,\, S_L) ,
\nonumber \\[-0.02cm]
T_{ij} 
& = \frac{1}{3} \delta_{ij} 
       - \text{Re} \, (\, E_i^{\, *} E_j \,) 
= \frac{1}{2} 
\left(
    \begin{array}{ccc}
     - \frac{2}{3} S_{LL} + S_{TT}^{xx}    & S_{TT}^{xy}  & S_{LT}^x  \\[+0.20cm]
     S_{TT}^{xy}  & - \frac{2}{3} S_{LL} - S_{TT}^{xx}    & S_{LT}^y  \\[+0.20cm]
     S_{LT}^x     &  S_{LT}^y              & \frac{4}{3} S_{LL}
    \end{array}
\right) .
\label{eqn:spin-1-vector-tensor-2}
\nonumber \\[-0.70cm]
\end{align}
%%%%%%%%%%%%%%%%%%%%%%%%%%%%%%%%%%%%%%%%%%%%%%%%%%%%%%%%%%%%%%%%%%
\begin{table}[t!]
\vspace{-0.37cm}
\footnotesize
\renewcommand{\arraystretch}{1.5}
\begin{center}
\begin{tabular}{|l|c|ccccc|} \hline
Polarizations & $\vec E$ & 
       $S_T^x$ & $S_T^y$ & $S_L$ & $S_{LL}$ & $S_{TT}^{xx}$ 
\\ \hline
%%%%%
Longitudinal $+z$ & $\frac{1}{\sqrt{2}} (-1,\, -i,\, 0)$ &
          0    &   0     &  $+$1 & $+\frac{1}{2}$ &   0  \\ \hline
Longitudinal $-z$ & $\frac{1}{\sqrt{2}} (+1,\, -i,\, 0)$ &
          0    &   0     &  $-$1 & $+\frac{1}{2}$ &   0  \\ \hline
%%%%%
Transverse $+x$ & $\frac{1}{\sqrt{2}} (0,\, -1,\, -i)$ &
         $+$1    &   0     &  0  & $-\frac{1}{4}$ & $+\frac{1}{2}$ \\ \hline
Transverse $-x$ & $\frac{1}{\sqrt{2}} (0,\, +1,\, -i)$ &
         $-1$    &   0     &  0  & $-\frac{1}{4}$ & $+\frac{1}{2}$ \\ \hline
%%%%%
Transverse $+y$ & $\frac{1}{\sqrt{2}} (-i,\, 0,\, -1)$ &
         0   &   $+$1      &  0  & $-\frac{1}{4}$ & $-\frac{1}{2}$ \\ \hline
Transverse $-y$ & $\frac{1}{\sqrt{2}} (-i,\, 0,\, +1)$ &
         0   &   $-1$     &  0  & $-\frac{1}{4}$ & $-\frac{1}{2}$ \\ \hline
%%%%%
Linear  $x$  &  $(1,\, 0,\, 0)$ &
         0   &   0      &  0  & $+\frac{1}{2}$ & $-1$ \\ \hline
Linear  $y$  &  $(0,\, 1,\, 0)$ &
         0   &   0     &  0  & $+\frac{1}{2}$ & $+1$ \\ \hline
%%%%%
\end{tabular}
\caption{Longitudinal, transverse, and linear polarizations
of the deuteron, polarization vectors, and parameters 
of spin vector and tensor \cite{ks-trans-g-2020,nica}.
The other parameters vanish in the considered polarizations
($S_{TT}^{xy} = S_{LT}^{x} = S_{LT}^{y}=0$).
The gluon transversity distribution is associated with 
the polarization parameter $S_{TT}^{xx}$.}
\label{table:polarizations}
\end{center}
\vspace{-0.8cm}
\end{table}
\normalsize
%%%%%%%%%%%%%%%%%%%%%%%%%%%%%%%%%%%%%%%%%%%%%%%%%%%%%%%%%%%%%%%%%%
% \ \vspace{-0.5cm} 

\noindent 
The parameters $S_{T}^x$, $S_{T}^y$, $S_L$, 
$S_{LL}$, $S_{TT}^{xx}$, $S_{TT}^{xy}$, $S_{LT}^x$, and $S_{LT}^y$
are assigned to express the vector and tensor polarizations
of the spin-1 deuteron.
The convariant forms $S^\mu$ and $T^{\mu\nu}$ are 
given in Ref.\,\cite{ks-trans-g-2020}.

The polarizations of the spin-1 deuteron are summarized in
Table \ref{table:polarizations} by showing 
the polarization $\vec E$ and the polarization parameters
for the longitudinal, transverse, and linear polarizations
of the deuteron.
For example, the longitudinal polarization contains
not only the longitudinal polarization parameter $S_L$
but also the tensor polarization $S_{LL}$.
The transverse polarization has
the transverse-polarization parameter $S_T^{x,y}$,
the tensor polarization $S_{LL}$, and
the linear polarization $S_{TT}^{xx}$.
The linear polarization has
both the tensor polarization $S_{LL}$ and
the linear polarization $S_{TT}^{xx}$.
Therefore, for extracting the gluon transversity $\Delta_T g$
associated with $S_{TT}^{xx}$, the linear polarization
asymmetry $E_x -E_y$ needs to be taken.
Alternatively, the transverse polarizations could be
used if the $S_T^{x,y}$ and $S_{LL}$ terms 
are removed by appropriate polarization combinations
or calculated by using corresponding PDFs
\cite{ks-trans-g-2020}.

\vspace{-0.40cm}
%%%%%%%%%%%%%%%%%%%%%%%%%%%%%%%%%%%%%%%%%%%%%%%%%%%%%%%%%%%%%%%%%%%%%%%%%%%%%%%%
\section{Gluon transversity}
\label{g-transversity}
\vspace{-0.20cm}

The longitudinally-polarized and transversity distributions
are defined for quarks and gluons 
by matrix elements of nonlocal operators as
\vspace{-0.10cm}
\begin{align}
\Delta q (x) & = \int \! \frac{d \xi^-}{4\pi} \, e^{i x p^+ \xi^-}
\langle \, p \, s_L  \left | \, \bar\psi (0) 
\gamma^+ \gamma_5 \psi (\xi)  \, \right | p \, s_L \, \rangle 
_{\xi^+=\vec\xi_\perp=0}  ,
\nonumber \\[-0.00cm]
\Delta_T q (x) & = \int \! \frac{d \xi^-}{4\pi} \, e^{i x p^+ \xi^-}
\langle \, p \, s_{T j} \left | \, \bar\psi (0)  
\, i \, \gamma_5 \, \sigma^{j +} 
 \psi (\xi) \, \right | p \, s_{Tj} \, \rangle 
_{\xi^+=\vec\xi_\perp=0} ,
\nonumber \\[-0.00cm]
\Delta_T g (x) 
& = \varepsilon_{TT,\alpha\beta}
\int \! \frac{d \xi^-}{2\pi} \, x p^+ \, e^{i x p^+ \xi^-} \,
\langle \, p \, E_{x} \left | \, A^{\alpha} (0) \, A^{\beta} (\xi)  
\right | p \, E_{x} \, \rangle 
_{\xi^+=\vec\xi_\perp=0}  ,
\label{eqn:pdf-definitions}
\nonumber \\[-0.70cm]
\end{align}
by using the lightcone coordinates.
Here, $\Delta q$ is the longitudinally-polarized
quark distribution function, 
and $\Delta_T q$ and $\Delta_T g$ are quark and gluon
transversity distributions.
The $p$ is the hadron momentum,
the $s_L$ and $s_{Tj}$ ($j=1$ or $2$) indicate longitudinal 
and transverse polarizations,
$E_x$ is the linear polarization, 
$\psi$ and $A^\mu$ are the quark and gluon fields,
and $x$ is the momentum fraction carried by a parton.
The gauge links for the color gauge invariance
are not explicitly written.
The transverse parameter $\varepsilon_{TT}^{\alpha\beta}$ 
is defined by $\varepsilon_{TT}^{11}=+1$ and $\varepsilon_{TT}^{22}=-1$.

The longitudinally-polarized quark distribution functions are given
by the difference between the quark distributions with spin parallel to
the hadron spin and the ones with antiparallel spin:
$\Delta q (x) = q_+ (x) - q_- (x)$ for the helicity operator
$\Pi_\parallel = \Sigma_\parallel/2$ with 
$\vec \Sigma = \gamma_5 \gamma_0 \vec \gamma$.
The transversity distribution is expressed as
$\Delta_T q (x) = q_\uparrow (x) - q_\downarrow (x)$,
where $\uparrow$ and $\downarrow$ indicate parallel and
anti-parallel quark polarizations
defined by the polarization operator 
$\tilde\Pi_\perp = \gamma_0 \Sigma_\perp/2$
\cite{ks-trans-g-2020}
in the transversely polarized hadron.
Equation (\ref{eqn:pdf-definitions}) indicates that the gluon
transversity is the distribution of linearly-polarized gluon
distribution difference in the linearly-polarized deuteron
\begin{align}
\Delta_T g (x) = g_{\hat x/\hat x} (x) - g_{\hat y/\hat x} (x) ,
\label{eqn:gluon-transversity-linear}
\nonumber \\[-0.70cm]
\end{align}
where $\hat y/\hat x$ indicate the gluon linear polarization
$\varepsilon_y$ in the deuteron with the polarization $E_x$.

%%%%%%%%%%%%%%%%%%%%%%%%%%%%%%%%%%%%%%%%%%%%%%%%%%%%%%%%%%%%%%%
\begin{wrapfigure}[8]{r}{0.37\textwidth}
   \vspace{-0.2cm}
   \hspace{-0.0cm}
\begin{minipage}[b]{0.37\textwidth}
\begin{center}
  \includegraphics[width=4.5cm]{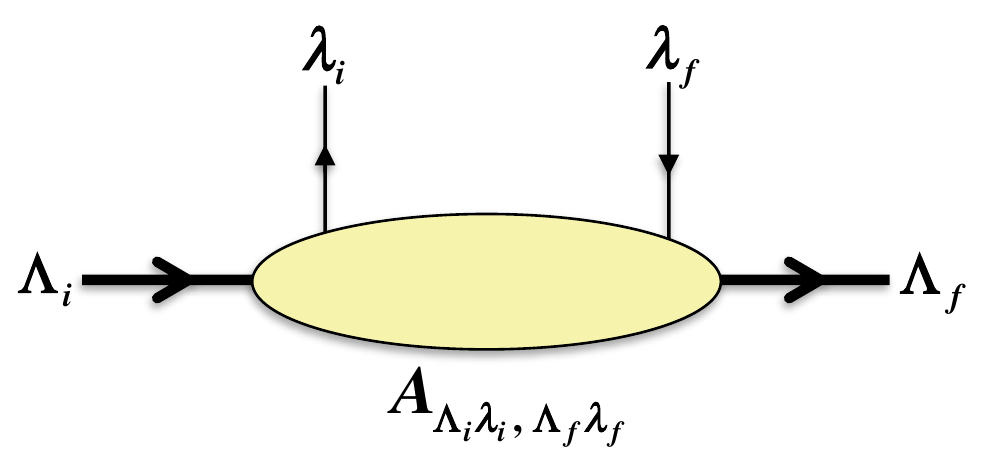}
\end{center}
\vspace{-0.70cm}
\caption{\hspace{-0.20cm} Parton-hadron forward scattering 
                          amplitudes.}
\label{fig:helicity-amp}
\end{minipage}
\end{wrapfigure}
%%%%%%%%%%%%%%%%%%%%%%%%%%%%%%%%%%%%%%%%%%%%%%%%%%%%%%%%%%%%%%%
Structure functions of hadrons are expressed by the imaginary part 
of parton-hadron forward scattering amplitudes
$A_{\Lambda_i \lambda_i ,\, \Lambda_f \lambda_f}$
in Fig.\,\ref{fig:helicity-amp},
where the initial and final hadron helicities are shown by
$\Lambda_i$ and $\Lambda_f$ and parton ones are by
$\lambda_i$ and $\lambda_f$.
These helicities satisfy the conservation relation
$\Lambda_i - \lambda_i = \Lambda_f - \lambda_f$.
The polarized quark and gluon distribution functions
defined in Eq.\,(\ref{eqn:pdf-definitions}) are expressed
by the helicity amplitudes as
\vspace{-0.10cm}
\begin{align}
\Delta q (x)    \sim \text{Im} \, (A_{++,\, ++} - A_{+-,\, +-}) \, , \ \ \ 
\Delta_T q (x)  \sim \text{Im} \, A_{++,\, - \hspace{0.03cm} -} \, , \ \ \ 
\Delta_T g (x)  \sim \text{Im} \, A_{++,\, - \hspace{0.03cm} -} \ .
\label{eqn:delta-deltaT-amplitudes}
\nonumber \\[-0.70cm]
\end{align}
Both quark and gluon transversity distributions are given
by the amplitudes with the helicity flips for both quark (gluon)
and hadron. Since the nucleon is spin 1/2, the quark-helicity flip
is possible and it has the quark transversity distribution.
However, the hadron spin should be one or larger for allowing
the helicity flip of two units for the gluon, so that
the gluon transversity does not exist in the nucleon.
As a spin-1 hadron or nucleus, the deuteron is most appropriate
for experimental investigations because it is a stable and simple
nucleus. These facts suggest a unique purpose
for investigating the gluon transversity of the deuteron.
The deuteron is a weak bound system of a proton and a neutron;
however, these nucleons cannot contribute to the gluon 
transversity directly. Therefore, if a finite distribution
is found experimentally, it sheds light on any new hadron physics
beyond the simple bound system of nucleons.

\vspace{-0.40cm}
%%%%%%%%%%%%%%%%%%%%%%%%%%%%%%%%%%%%%%%%%%%%%%%%%%%%%%%%%%%%%%%%%%%%%%%%%%%%%%%%
\section{Proton-deuteron Drell-Yan process and gluon transversity}
\label{pd-drell-yan}
\vspace{-0.30cm}

%%%%%%%%%%%%%%%%%%%%%%%%%%%%%%%%%%%%%%%%%%%%%%%%%%%%%%%%%%%%%%%
\begin{wrapfigure}[17]{r}{0.33\textwidth}
   \vspace{-0.3cm}
   \hspace{-0.0cm}
\begin{minipage}[b]{0.33\textwidth}
\begin{center}
  \hspace{-0.30cm}
  \includegraphics[width=4.0cm]{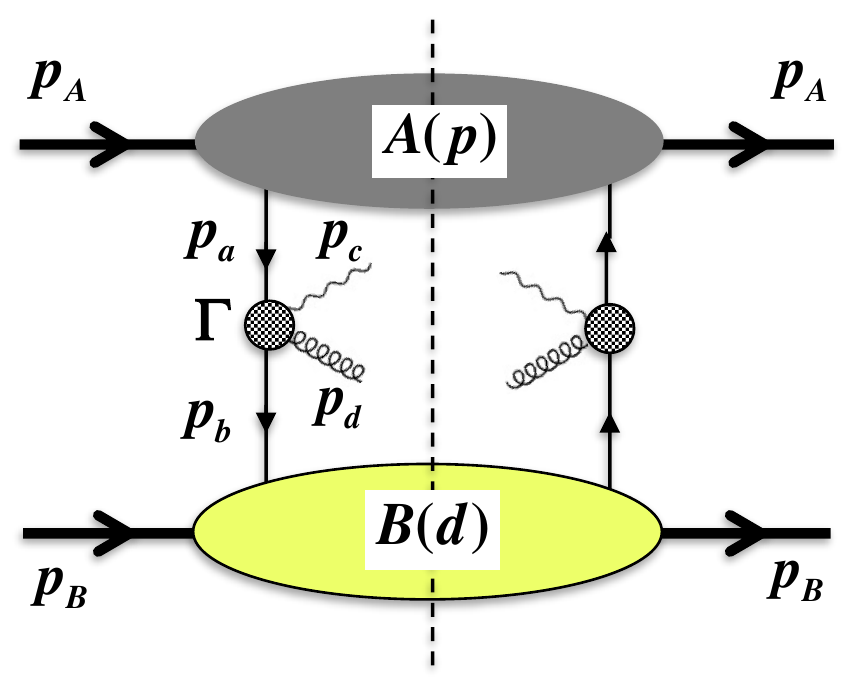}
\end{center}
\vspace{-0.80cm}
\caption{\hspace{-0.20cm} $q\bar q \to \gamma^* g$ subprocess.}
\label{fig:qqbar-dy}
\end{minipage}\\
\begin{minipage}[b]{0.33\textwidth}
\vspace{0.20cm}
\begin{center}
  \includegraphics[width=4.0cm]{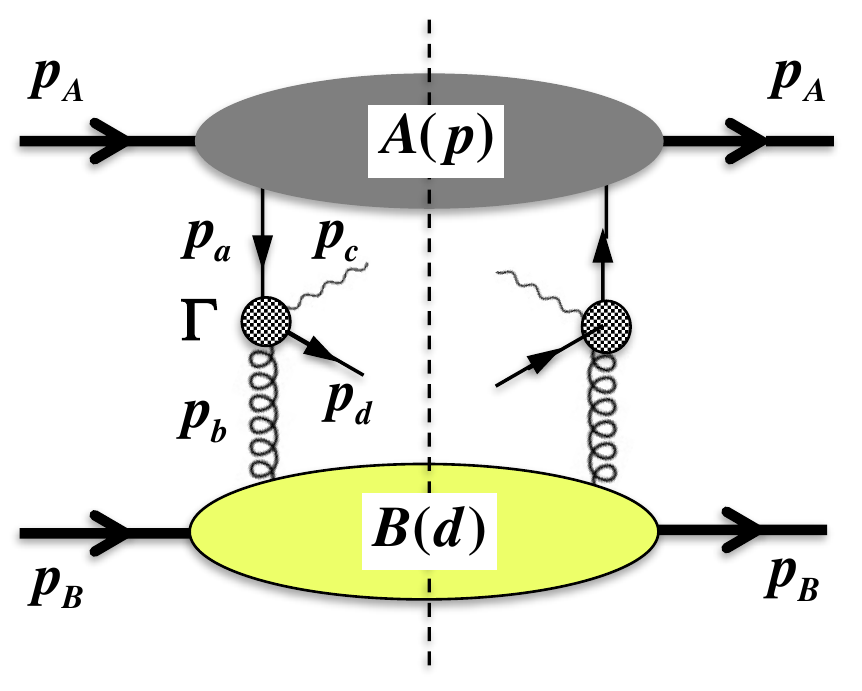}
\end{center}
\vspace{-0.80cm}
\caption{\hspace{-0.20cm} $qg \to \gamma^* q$ subprocess.}
\label{fig:qg-dy}
\end{minipage}
\end{wrapfigure}
%%%%%%%%%%%%%%%%%%%%%%%%%%%%%%%%%%%%%%%%%%%%%%%%%%%%%%%%%%%%%%%

The purpose of our research is to propose a possibility
to investigate the gluon transversity at hadron accelerator
facilities. As an example, we made the theoretical formalism
for the proton-deuteron Drell-Yan process $p+d \to \mu^+\mu^-+X$
with the linearly-polarized deuteron.
There are many subprocesses which contribute to the Drell-Yan 
cross section as typically shown in Figs.\,\ref{fig:qqbar-dy}
and \ref{fig:qg-dy} for the subprocesses 
$q\bar q \to \gamma^* g$ and $qg \to \gamma^* q$.
%%%%%
The upper proton blob and the lower deuteron blob
indicate the collinear correlation functions,
which are expressed by various parton distribution functions
(PDFs). The proton is not polarized, so that only the unpolarized
PDFs of the proton are needed for estimating the cross section.
On the other hand, we consider the linear polarizations
$E_x$ and $E_y$ for the deuteron, and their asymmetry
$d\sigma (E_x)-d\sigma (E_y)$ is calculated.
We take the $z$ axis as the proton momentum direction.
The differential cross section is calculated for 
the variables $\tau=M_{\mu\mu}^2/s=Q^2/s$,
the dimuon (virtual photon) transverse momentum squared $\vec q_T^2$, 
its azimuthal angle $\phi$,
and rapidity $y$ in the center-of-mass frame.
Here, $M_{\mu\mu}$ is the dimuon mass,
and  $s$ is the center-of-mass energy squared.
For the linear polarization asymmetry $E_x-E_y$, we obtained
the leading cross-section expression 
by integrating over the momentum fraction $x_a$ 
for partons in the proton as
\vspace{-0.10cm}
\begin{align}
\frac{ d \sigma_{pd \to \mu^+ \mu^- X} }{d\tau \, d \vec q_T^2 \, d\phi \, dy}
(E_x-E_y )
& = - \frac{\alpha^2 \, \alpha_s \, C_F \, q_T^2}{6\pi s^3} \cos (2\phi) 
\int_{\text{min}(x_a)}^1 dx_a 
 \frac{1} { (x_a x_b)^2 \, (x_a -x_1) (\tau -x_a x_2 )^2}
\nonumber \\
& \ \hspace{3.0cm}
 \times 
 \sum_{q}  e_q^2 \, x_a \!
 \left[ \, q_A (x_a) + \bar q_A (x_a) \, \right ]
  x_b \Delta_T g_B (x_b) ,
\label{eqn:cross-5}
\nonumber \\[-0.90cm]
\end{align}
where $\alpha$ is the fine structure constant,
$\alpha_s$ is the QCD running coupling constant,
$C_F$ is the color factor $C_F=(N_c^2-1)/(2N_c)$ with $N_c=3$,
and $e_q$ is the quark charge.
The momentum fraction $x_b$ is given by
$x_b=(x_a x_2 -\tau)/(x_a-x_1)$, and the kinematical minimum 
of $x_a$ is $\text{min}(x_a)=(x_1-\tau)/(1-x_2)$
with $x_1 = e^y \sqrt{(Q^2+\vec q_T^2)/s}$
and $x_2 = e^{-y} \sqrt{(Q^2+\vec q_T^2)/s}$.
The $q_A (x_a)$ and $\bar q_A (x_a)$ 
are quark and antiquark distribution functions in the proton,
and $\Delta_T g (x_b)$ is the gluon transversity.

\begin{figure}[h!]
\vspace{-0.20cm}
\hspace{0.3cm}
\begin{minipage}[c]{0.30\textwidth}
     \includegraphics[width=6.9cm]{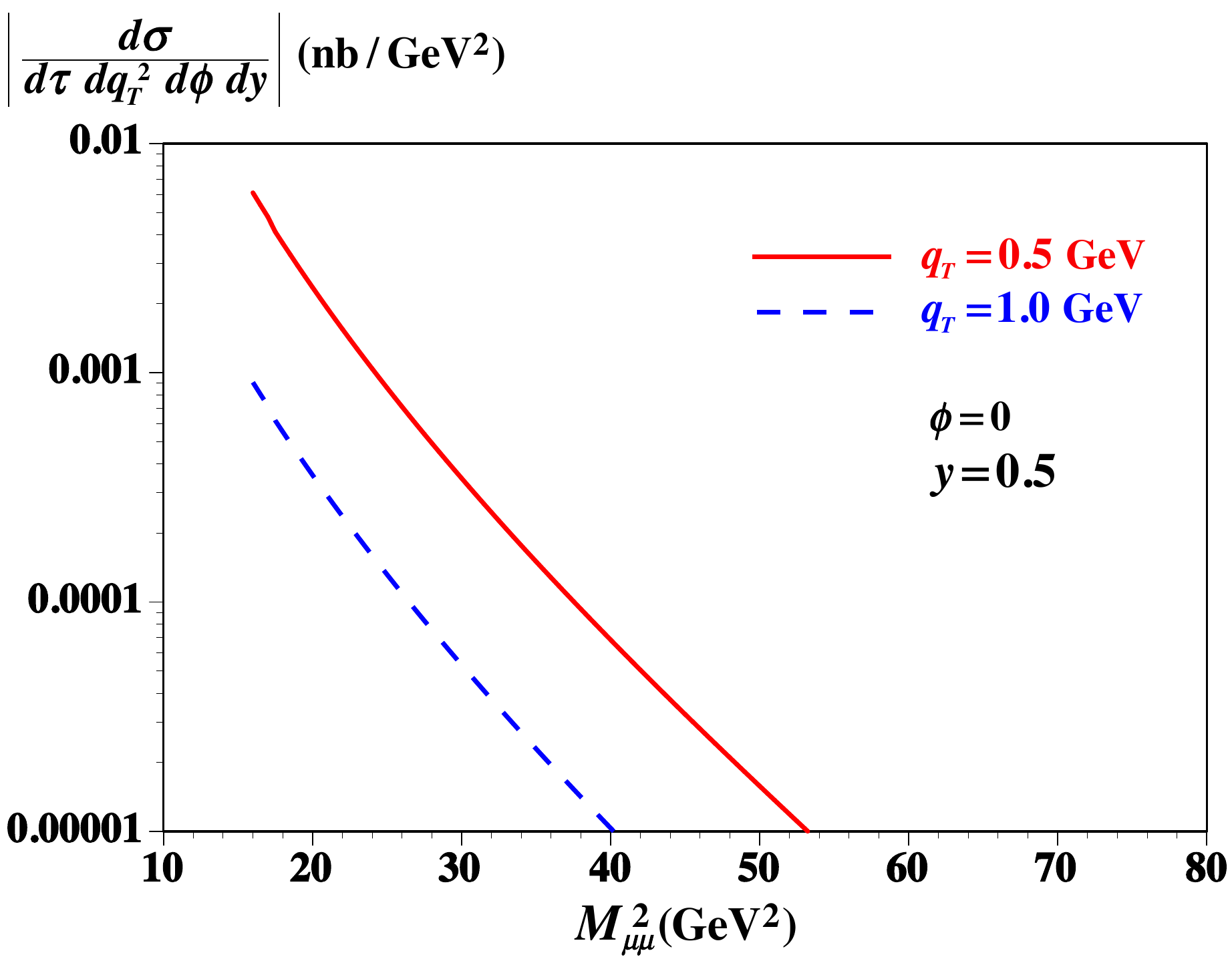}
\vspace{-0.70cm}
\caption{
Proton-deuteron Drell-Yan cross section \\
for the linear-polarization asymmetry $E_x-E_y$.
}
\label{fig:cross-ex-ey}
\end{minipage} 
\hspace{+4.0cm}
\begin{minipage}[c]{0.40\textwidth}
    \vspace{-0.00cm}
     \hspace{-0.50cm}
     \includegraphics[width=6.53cm]{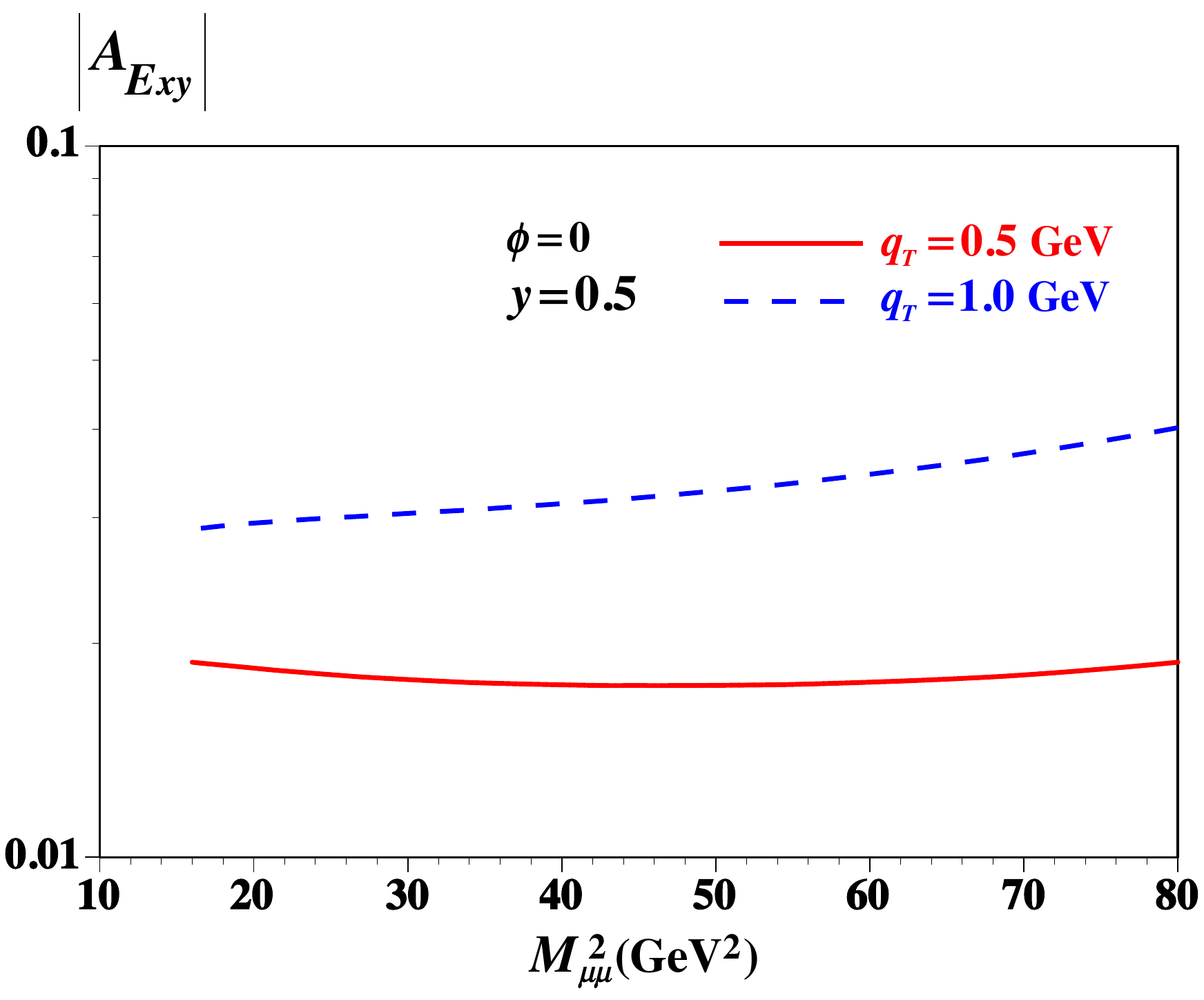}
\vspace{-0.70cm}
\caption{Spin asymmetry \\ \ \ $d\sigma (E_x-E_y)/d\sigma (E_x+E_y)$.}
\label{fig:asym-ex-ey}
\end{minipage}
\end{figure}

The absolute value of the cross section in Eq.\,(\ref{eqn:cross-5})
is shown in Fig.\,\ref{fig:cross-ex-ey} 
for the Fermilab kinematics with $p_p = 120$ GeV
by taking $\phi=0$, $y=0.5$, and $q_T=0.5$ or 1.0 GeV
as the function of the dimuon-mass squared $M_{\mu\mu}^2$.
Here, the CTEQ14 PDFs are used for the unpolarized PDFs of the proton, 
and the NNPDF1.1 is used for the gluon transversity by boldly
assuming that the longitudinally-polarized gluon distribution is
equal to the gluon transversity. 
Because of this bold assumption, the cross section and
the following spin asymmetry could be overestimated.
The cross section $d\sigma (E_x+E_y)$ can be calculated
by using the unpolarized PDFs of the proton and deuteron
with the assumption of neglecting the small tensor-polarized
PDF part, and the calculated spin assymmetry 
$d\sigma (E_x-E_y)/d\sigma (E_x+E_y)$
is shown in Fig.\,\ref{fig:asym-ex-ey}.
The asymmetries are typically a few percent.
If a finite value is found experimentally,
it could create a new field of hadron and nuclear physics
beyond the simple bound system of nucleons. 
Therefore, we hope that such an experiment is realized in future.
In fact, this experiment will be proposed within 
the Fermilab-E1039 experiment \cite{Fermilab-dy},
and a similar $J/\psi$ experiment should be possible
in the NICA project.

%%%%%%%%%%%%%%%%%%%%%%%%%%%% figure %%%%%%%%%%%%%%%%%%%%%%%%%%%%
%%%%%%%%%%%%%%%%%%%%%%%%%%%% figure %%%%%%%%%%%%%%%%%%%%%%%%%%%%

\vspace{-0.95cm}
%%%%%%%%%%%%%%%%%%%%%%%%%%%%%%%%%%%%%%%%%%%%%%%%%%%%%%%%%%%%%%%%%%%%%%%%%%%%%%%%
\section{Conclusion}
\label{conclusion}
\vspace{-0.35cm}

The gluon transversity has not been measured yet experimentally.
If a finite distributions is found, it indicates the existence of
new hadronic physics. In this work, we showed the general formalism
to investigate the gluon transversity at hadron accelerator facilities
by taking the Drell-Yan process as an example.
In the 2020's, we expect that the gluon transversity will be
measured at JLab, Fermilab, NICA, and other accelerator facilities,
so that it could become an exciting field of hadron physics
in the near future.

\vspace{-0.55cm}
%%%%%%%%%%%%%%%%%%%%%%%%%%%%%%%%%%%%%%%%%%%%%%%%%%%%%%%%%%%%%%%%%%%%%%%%%%%%%%%%
\section*{Acknowledgements}
\vspace{-0.35cm}

S. Kumano was partially supported by 
Japan Society for the Promotion of Science (JSPS) Grants-in-Aid 
for Scientific Research (KAKENHI) Grant Number 19K03830.
Qin-Tao Song was supported by the National Natural Science Foundation 
of China under Grant Number 12005191 and the Academic Improvement Project 
of Zhengzhou University.

\vspace{-0.55cm}
%%%%%%%%%%%%%%%%%%%%%%%%%%%%%%%%%%%%%%%%%%%%%%%%%%%%%%%%%%%%%%%%%%%%%%%%%%%%%%%%

%%%%%%%%%%%%%%%%%%%%%%%%%%%%%%%%%%%%%%%%%%%%%%%%%%%%%%%%%%%%%%%%%%%%%%%%%%%%%%%%

\end{document}